# Multi-Channel Non-Abelian Edge Pumping via Multi-Degeneracy Encircling

Qingyang Mo[1], Jiacheng Sun[2], Zhongfu Li[1], Jiajun Li[1], Wange Song[1,2*], Tao Li[2†], and Shuang Zhang[1,3,4,5,6‡]

[1]*New Cornerstone Science Laboratory, Department of Physics, The University of Hong Kong, 999077 Hong Kong, China*

[2]*National Laboratory of Solid State Microstructures, Key Laboratory of Intelligent Optical Sensing and Manipulations, Jiangsu Key Laboratory of Artificial Functional Materials, College of Engineering and Applied Sciences, Nanjing University, Nanjing, 210093, China*

[3]*State Key Laboratory of Optical Quantum Materials, The University of Hong Kong, 999077 Hong Kong, China*

[4]*Department of Electrical and Electronic Engineering, The University of Hong Kong, 999077 Hong Kong, China*

[5]*Quantum Science Center of Guangdong-Hong Kong-Macao Great Bay Area, 3 Binlang Road, Shenzhen, China*

[6]*Materials Innovation Institute for Life Sciences and Energy (MILES), HKU-SIRI, Shenzhen, People's Republic of China*

Non-Abelian operations enable sequence-dependent control of light, yet experimental realizations in wave systems have remained confined to local bulk or same-boundary modes and have often required fragile gap-closing degeneracies. Conventional Thouless pumps, meanwhile, support only two-channel edge transfer across a single gap. Here we demonstrate multi-channel non-Abelian edge pumping on a silicon photonic chip by adiabatically encircling multiple band degeneracies in a multi-gap lattice. In a minimal three-band "band-triangle", two opposite-charge Weyl points and one zero-charge degeneracy are linked by global band connectivity, yielding both opposite-boundary and same-boundary edge channels. Sequencing these channels produces non-Abelian braiding of three topologically protected edge modes that reside on distinct physical boundaries. The process remains fully gapped and is protected by topological winding rather than by maintained spectral degeneracies. The same band-triangle principle extends systematically to an arbitrary number of channels, establishing a scalable route to collective, sequence-controlled routing of topological states in integrated photonic circuits.

The non-commutative nature of non-Abelian operations offers a fundamentally new paradigm for information processing, where the sequence of operations determines the outcome. Originating in Yang–Mills gauge theory [1,2] and extending to non-Abelian anyons in topological quantum computation [3–7], this concept has recently been implemented in classical wave systems (e.g., acoustic and photonic systems), opening avenues for sequence-dependent control of light in photonic circuits [8–19]. Two main approaches have dominated experimental realizations of non-Abelian effects in photonic waveguide systems. The first draws inspiration from non-Abelian anyons, utilizing three or more degenerate zero modes to achieve braiding-like permutations [11–13]. The second is rooted in non-Abelian gauge theory: adiabatic paths traversing band degeneracies accumulate Wilczek–Zee phases, manifesting as non-Abelian Thouless pumping [14–16]. However, the non-commutative dynamics in both paradigms act upon degenerate bulk bands or tightly localized modes, confining all reported results to local, short-range behavior.

Thouless pumping provides a robust mechanism for long-range edge-to-edge transport [20–22]. In the Rice–Mele model, cyclic modulation of the dimerization and a staggered mass produces a Weyl point (WP) in parameter space. An adiabatic path that encircles this point pumps an edge mode from one boundary to the opposite boundary. This mechanism has been realized in cold-atom [23,24], photonic [25–32] and acoustic [33–36] systems. Yet the underlying single-gap structure supports only a single pair of opposite-boundary edge modes. Multi-band pumping that addresses multiple edge channels, and thereby enables non-Abelian operations among them, has remained out of reach [37–46].

Here we overcome both limitations by demonstrating multi-channel non-Abelian edge pumping on a silicon photonic chip. We introduce a minimal three-band "band-triangle" that hosts two WPs of opposite charge and one zero-charge degeneracy linked by global band connectivity. Encircling the WPs activates opposite-boundary channels; encircling the zero-charge degeneracy activates a same-boundary channel. Sequencing these elementary channels produces non-Abelian braiding of topologically protected edge states that reside on distinct physical boundaries. The construction extends systematically to an arbitrary number of channels by stacking additional band-triangles. We experimentally realize the minimal ($n = 3$) case, observe both classes of pumping channels, and demonstrate sequence-dependent transfer among the three edge modes. The process remains fully gapped and is protected by topological winding around the degeneracies, establishing a scalable route to collective, sequence-controlled routing of topological edge states in integrated photonic circuits.

*Minimal "band-triangle" model.* We begin with the minimal three-site lattice model supporting three-channel edge pumping—the band-triangle. The tight-binding Hamiltonian on a one-dimensional rhombic lattice with three sublattices A, B and C (Fig. 1(a)) reads

$$\hat{H}(\phi) = -\sum_i \left[\kappa_1(\phi)\left(\hat{a}_i^\dagger \hat{c}_{i-1} + \hat{b}_i^\dagger \hat{c}_{i-1}\right) + \kappa_2(\phi)\left(\hat{a}_i^\dagger \hat{c}_i + \hat{b}_i^\dagger \hat{c}_i\right) + h.c.\right] + \sum_i \left[m(\phi)\left(-\hat{a}_i^\dagger \hat{a}_i + \hat{b}_i^\dagger \hat{b}_i\right) + m_c \hat{c}_i^\dagger \hat{c}_i\right], \quad (1)$$

where $\hat{a}_i^\dagger$, $\hat{b}_i^\dagger$, and $\hat{c}_i^\dagger$ ($\hat{a}_i$, $\hat{b}_i$, and $\hat{c}_i$) are the creation (annihilation) operators acting on the A, B, and C sites of the $i$-th unit cell. The inter-(intra) couplings are modulated as $\kappa_{1,2}(\phi) = \kappa_0 \pm \Delta(\phi)/2$, while onsite potentials are applied as $m_{A,B} = \mp m(\phi)$ and $m_C = -\kappa_0$, where $\phi$ is the adiabatic pumping parameter tracing closed loops in the $(m, \Delta)$ plane. As shown in Fig. 1(b), the energy spectrum versus $m$ at $(\Delta, k) = (0, \pi)$ exhibits three degeneracies that form a closed triangle: A pair of WPs with the charge of $\pm 1$ (red/blue) and a zero-charge degeneracy (yellow). The effective Hamiltonians around these degeneracies are presented in Supplementary Materials (SM), Section 1.

We first examine the adiabatic paths that enclose only one of the charged WPs. When the trajectory in the $(m, \Delta)$ plane encircles solely the negative-charge WP (blue loop in Fig. 1(c)), the edge spectrum evolves as shown in Fig. 1(d). The grey regions denote bulk bands, while the colored branches are edge modes localized on the A (blue), B (red) and C (purple) sublattices. For the middle band, the state starts from the C-localized edge mode and continuously transfers to the A-localized edge mode, realizing an opposite-boundary pump with Chern number $C = -1$, consistent with the topological charge of the encircled WP. Conversely, a path that encloses only the positive-charge WP (red loop) produces the exchange between the B- and C-localized edge modes with $C = 1$ for the middle band (Fig. 1(e)). In both cases the third edge mode remains spectrally detached throughout the cycle.

These two opposite-boundary processes already involve all three edge modes. Therefore, the spectral continuity requires that the A- and B-localized edge modes be connected to each other after the A-C and C-B exchanges. The only available locus for this connection is the zero-charge degeneracy (yellow point). Consequently, an adiabatic path that encloses the entire band-triangle must produce a same-boundary exchange between the A and B edge modes. This expectation is confirmed in Fig. 1(f): when all three degeneracies are encircled (yellow loop in Fig. 1(c)), the edge spectrum exhibits an A–B transfer on the same boundary. The Chern number for this same-boundary channel vanishes, consistent with the zero charge of the yellow degeneracy.

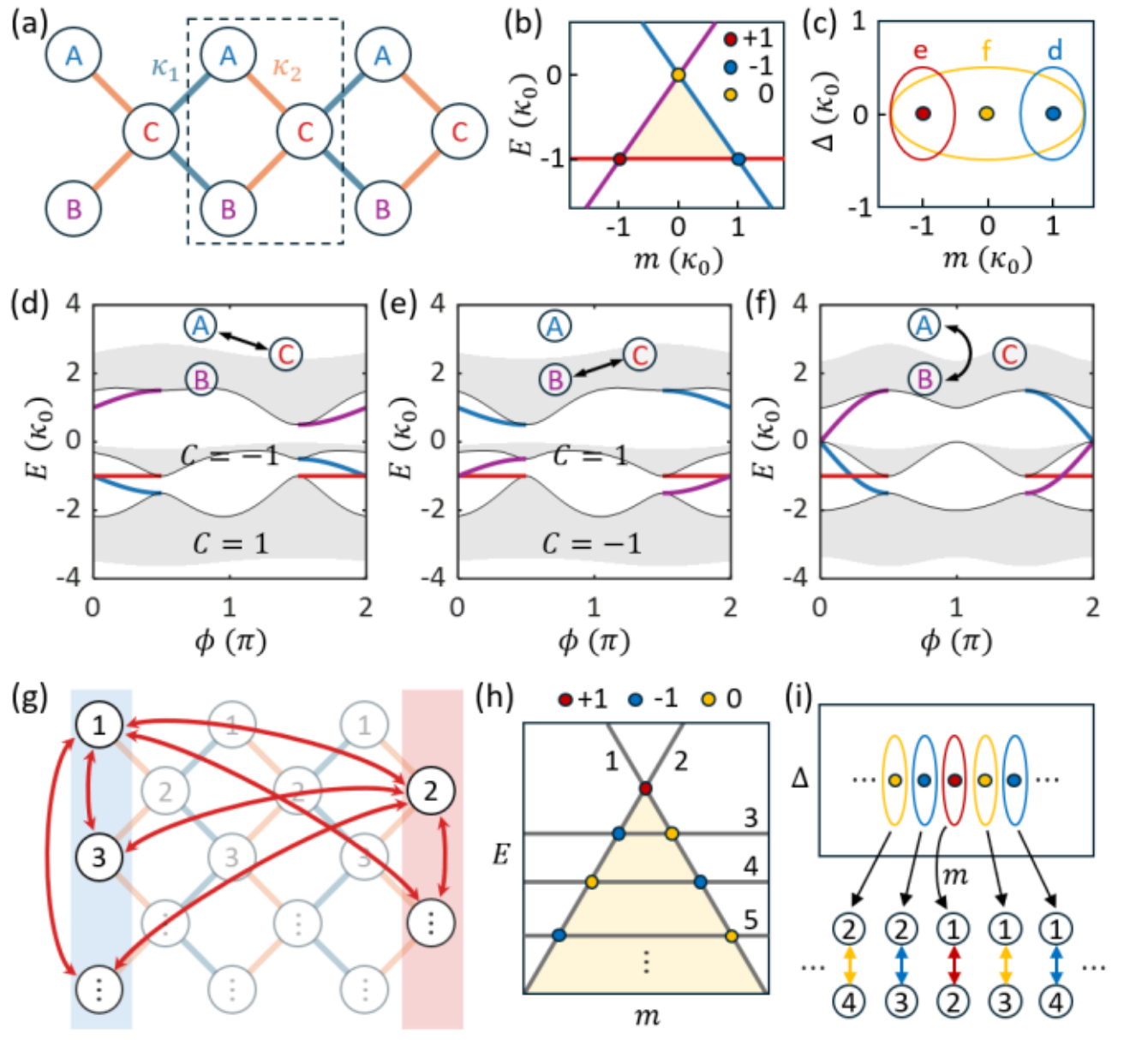

FIG. 1. Multi-channel edge pumping via the band-triangle. (a) Three-site rhombic lattice. The on-site potentials on $\{A, B, C\}$ sites take the form of $\{-m(\phi), m(\phi), -\kappa_0\}$. (b) Energy spectrum versus $m$ at $(\Delta, k) = (0, \pi)$, forming a band-triangle with charges +1 (red), -1 (blue), and 0 (yellow). (c) Adiabatic paths in the $(m, \Delta)$ plane enclosing different degeneracies. (d–f) Calculated energy spectra for the paths in (c). Grey regions denote bulk bands, and colored branches are edge modes localized on A (blue), B (red) and C (purple) sublattices. $m = \kappa_0 + 0.5\kappa_0 \sin\phi$ for (d), $m = -\kappa_0 + 0.5\kappa_0 \sin\phi$ for (e), and $m = 1.5\kappa_0 \sin\phi$ for (f). $\Delta = \kappa_0 \cos\phi$ for three paths. (g) Extensible $n$-site lattice; odd (even) sites form the left (right) boundary. (h) Stacked band-triangles in the multi-band spectrum. (i) Each degeneracy-encircling loop maps onto a specific edge-mode exchange.

This band-triangle principle can be extended immediately to an arbitrary number of channels. Consider an extensible one-dimensional lattice with $n$ sites per unit cell arranged in a rhombic geometry (Fig. 1(g)). Odd-indexed ($i = 1, 3, 5, \ldots$) sites form one boundary (blue shading) and even-indexed ($i = 2, 4, 6, \ldots$) sites the opposite boundary (red shading). Staggered potentials $\mp m(\phi)$ are assigned to a chosen pair (for example, sublattices 1 and 2), while the remaining sublattices receive fixed gradient potentials (for example, $\{-m_0, -2m_0, \ldots\}$). The resulting spectrum contains $(2n-3)$ degeneracies organized into stacked band-triangles (Fig. 1(h)): charged WPs $(\pm 1)$ mediate opposite-boundary exchange, and zero-charge degeneracies mediate same-boundary exchange.

Each degeneracy-encircling loop in the $(m, \Delta)$ plane maps onto a specific edge-mode transfer (Fig. 1(i)). Encircling a single charged WP pumps an edge mode between opposite boundaries (e.g., $1 \leftrightarrow 2$, $2 \leftrightarrow 3$), while encircling a zero-charge degeneracy pumps between sites on the same boundary (e.g., $1 \leftrightarrow 3$, $2 \leftrightarrow 4$). Consequently, an edge mode initialized on site 1 or 2 can be directed to any other edge site by enclosing the appropriate degeneracy. Adding further sites simply introduces additional triangles and channels, rendering the construction systematically scalable. For $n = 4$, every single-degeneracy path produces a well-defined edge-to-edge exchange while the other edge modes remain detached (see SM, Section 2), as predicted by the stacked band-triangle picture.

*Photonic chip design.* We implement the three-site band-triangle model on a silicon photonic chip using a custom multi-layer fabrication process on a silicon-on-sapphire substrate. Fig. 2(a) illustrates the schematic cross-section of the tri-layer waveguide array, consisting of waveguides A (top layer), B (bottom layer), and C (middle layer) corresponding to sublattices in tight-binding model, with the propagation direction $z$ serving as the adiabatic pumping parameter $\phi$. Inter-layer evanescent couplings $\kappa_1(\phi)$ and $\kappa_2(\phi)$ are tuned by the horizontal separations $d_1$ (inter-cell) and $d_2$ (intra-cell), while staggered onsite potentials $\mp m(\phi)$ on A/B sublattices and constant $m_C$ on C are achieved by varying waveguide widths $w_A$, $w_B$, and $w_C$ along $z$. More details about structure parameters are presented in SM, Section 3. To experimentally observe the edge dynamics, we fabricated a waveguide array consisting of three unit cells. Fig. 2(b) provides an overall schematic of the device layout. Grating couplers at the input facet enable selective excitation of edge modes (A1, B1, and C3), while a fan-out structure at the output facet routes the light from waveguides to individual output ports (A1, B1, C1, …, A3, B3, C3), allowing direct measurement of the field intensities after the pumping region. The inset shows a scanning electron microscope (SEM) image at the midpoint ($z = L/2$) of the pumping area, demonstrating the high-quality fabrication of the tri-layer waveguide array.

Two representative adiabatic paths are implemented by sinusoidal modulation of the structural parameters along $z$ (see details in SM, Section 3). Fig. 2(c) shows the parametric profiles of $\kappa_1(\phi)$, $\kappa_2(\phi)$, and $m(\phi)$ for a path that encloses only a single negative-charge WP and activates the opposite-boundary (A-C) channel. The simulated spectrum of effective propagation constants versus $\phi$ (Fig. 2(d)) indicates an edge-mode exchange in the lower band gap (red lines), while the third edge mode remains detached (yellow line). Fig. 2(e) further plots the simulated field distributions for three edge states marked by i-iii in Fig. 2(d), showing a good localization at the edge and a clear mode-exchange between A1 and C3 edge modes. This strong localization ensures that residual coupling between opposite-boundary modes remains negligible already at three cells. Pumping-fidelity calculations for different cell numbers further confirm that high-fidelity edge-to-edge transfer is achieved for all lattices with the cell number $N > 2$ (see SM, Section 4). The second path, whose parametric profiles are shown in Fig. 2(f), encloses the entire band-triangle and activates the same-boundary (A-B) channel. The simulated band spectrum shows a pair of exchanged edge bands connecting the top and bottom bulk bands (Fig. 2(g)), matching well with the tight-binding prediction (Fig. 1(f)). The simulated field distributions for edge states again certify the exchange between highly localized A1 and B1 edge modes (Fig. 2(h)).

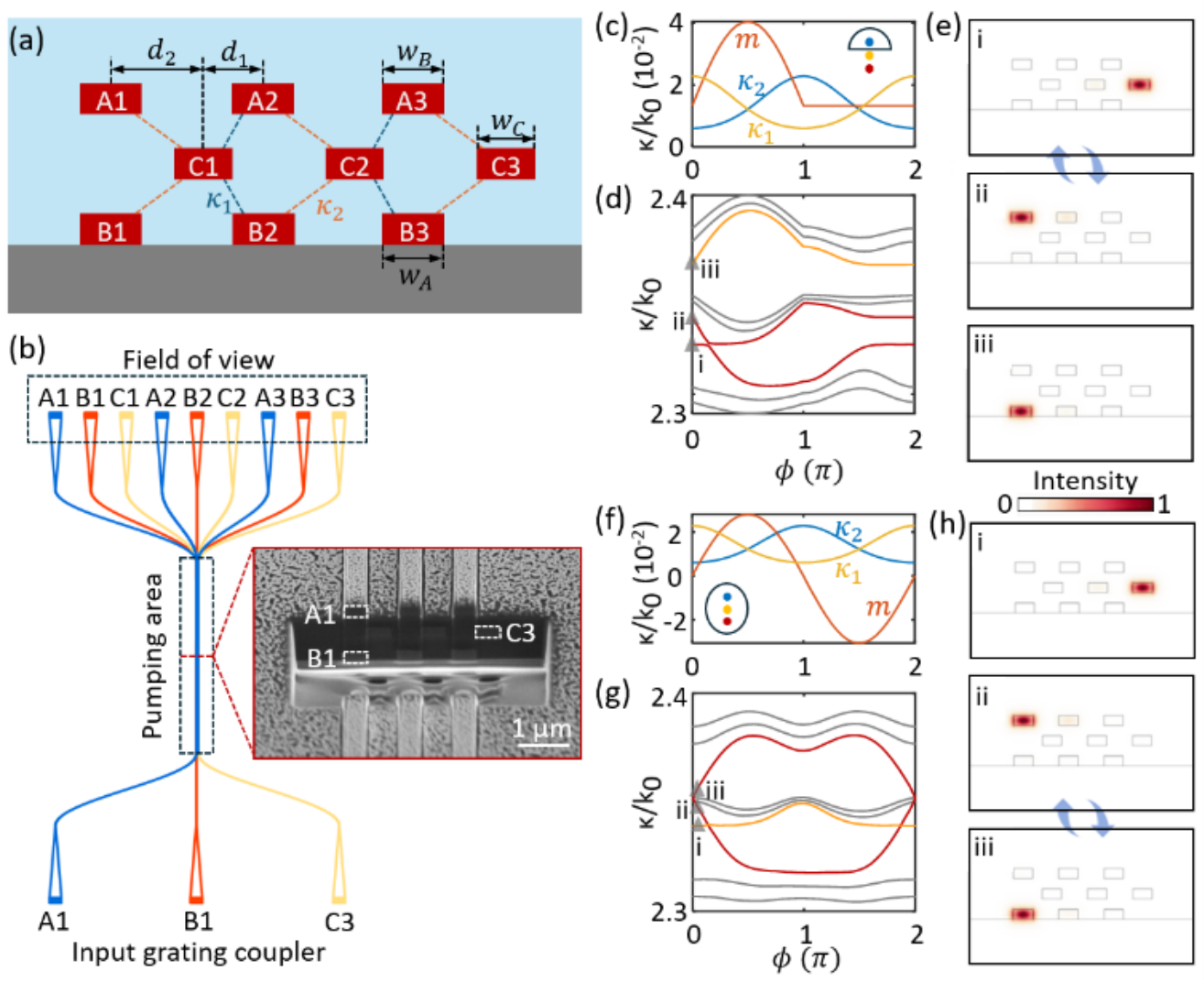

FIG. 2. Photonic chip design and adiabatic paths. (a) Schematic of the cross-section geometry in the waveguide array. (b) Schematic of experimental sample design. Inset: Scanning electron microscope (SEM) image showing the cross section of the waveguide array at $z = L/2$, where $L$ is the pumping length. (c-e) The pumping path encloses a negative-charge WP, activating the opposite-boundary (A–C) channel. (c) Profiles of $\kappa_1(\phi)$, $\kappa_2(\phi)$ and $m(\phi)$. (d) Simulated spectrum of effective propagation constants versus $\phi$. (e) Simulated field distributions for three edge states marked by i-iii in (d). (f-h) Corresponding results for the pumping path that encloses the whole band triangle,

activating the same-boundary (A–B) channel.

*Experimental observations in on-chip waveguides.* We first characterize the two elementary adiabatic paths separately. For the path that encloses only a single negative-charge WP (opposite-boundary channel), light is selectively launched into the edge modes A1, B1 and C3. After propagation through the modulated region, the measured output intensity distributions (bottom row of Fig. 3(c)) show a clear transfer of localization between A1 and C3 across opposite boundaries, while the B1 waveguide remains largely decoupled (Fig. 3(a)). Restricting the output to these three working edge ports and renormalizing column-wise gives a permutation matrix for the A1 ↔ C3 exchange (Fig. 3(b)). These results, together with the full-wave simulations (top row of Fig. 3(c)), confirm the opposite-boundary A–C Thouless pump associated with the charged WP.

For the path that encloses the entire band-triangle (same-boundary channel), the same input configuration is used. The measured output intensity distributions (bottom row of Fig. 3(f)) now exhibit an exchange of localization between A1 and B1 on the left boundary, while light injected into C3 remains largely confined to the C3 output port with negligible crosstalk (Fig. 3(d)). The measured 3×3 map in Fig. 3(e) records this same-boundary permutation. The observations demonstrate the same-boundary A–B pumping channel enforced by the zero-charge degeneracy, in good agreement with the full-wave simulations (top row of Fig. 3(f)).

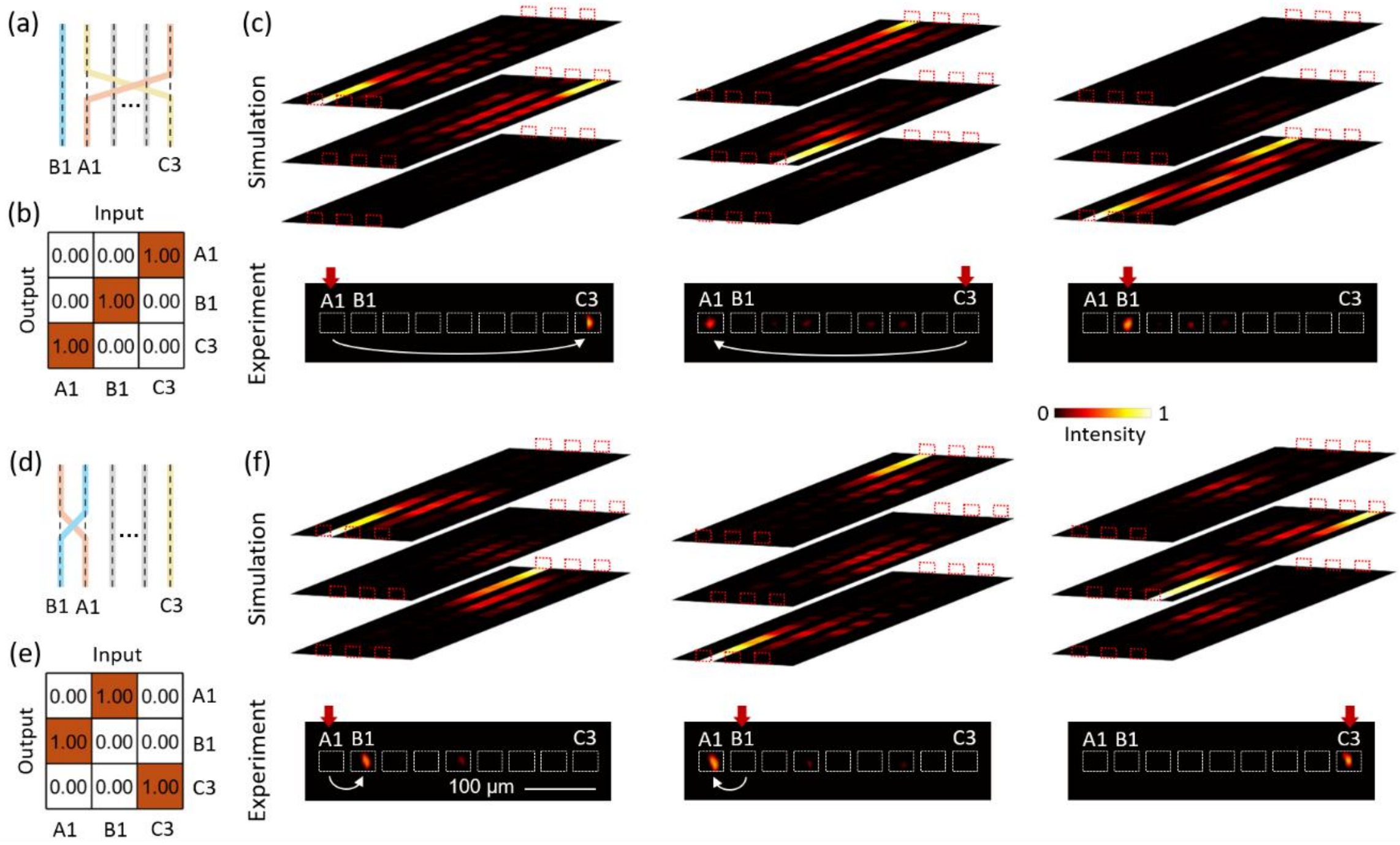

Fig. 3. Experimental demonstration of the two elementary pumping channels. (a-c) Opposite-boundary channel realized by a path that encloses a single negative-charge Weyl point. (a) Schematic of the A1 ↔ C3 exchange. (b) Measured 3×3 map among the working edge ports A1, B1 and C3. (c) Full-wave simulations (top) and measured output intensity distributions (bottom) for inputs at A1, B1 and C3 (red arrows). (d-f) Corresponding results for the same-boundary channel realized by a path that encloses the entire band-triangle. The experimental pumping length $L = 500\mu\text{m}$.

Together, the two elementary paths realize the complete set of three-channel operations predicted by the band-triangle: an edge mode can be directed either to the opposite boundary or to the other site on the same boundary, depending on which degeneracies are enclosed.

To demonstrate the non-commutative character of the multi-channel pumping, we concatenate the two elementary paths in alternating order, forming composite adiabatic cycles. When the same-boundary channel is executed first and then followed by the opposite-boundary channel, the three edge modes undergo the cyclic permutation A1 → B1 → C3 → A1 (see Fig. 4(a-c)). Conversely, when the opposite-boundary channel is applied first and then followed by the same-boundary channel, the distinct cycle A1 → C3 → B1 → A1 is obtained (see Fig. 4(d-f)).

The measured output intensity distributions for the two sequences are shown in bottom rows of Figs. 4(c) and 4(f), respectively. For each input (A1, B1 or C3), the dominant intensity peak appears at the port predicted by the corresponding cyclic permutation. Residual light appears in bulk waveguides and reduces the overall conversion efficiency (see SM, Section 7), but the assignment among the three working edge ports remains of high purity. The corresponding 3×3 maps (Figs. 4(b) and 4(e)) have target-port occupancies of unity within the measurement resolution for one sequence and above 0.9 for the other. Full-wave simulations (top rows of Figs. 4(c) and 4(f)) reproduce the sequence-specific intensity patterns. The clear difference between the two composite paths constitutes direct experimental evidence that the edge-state transfer is non-Abelian: the final destination of a given input depends on the order in which the opposite-boundary and same-boundary channels are applied.

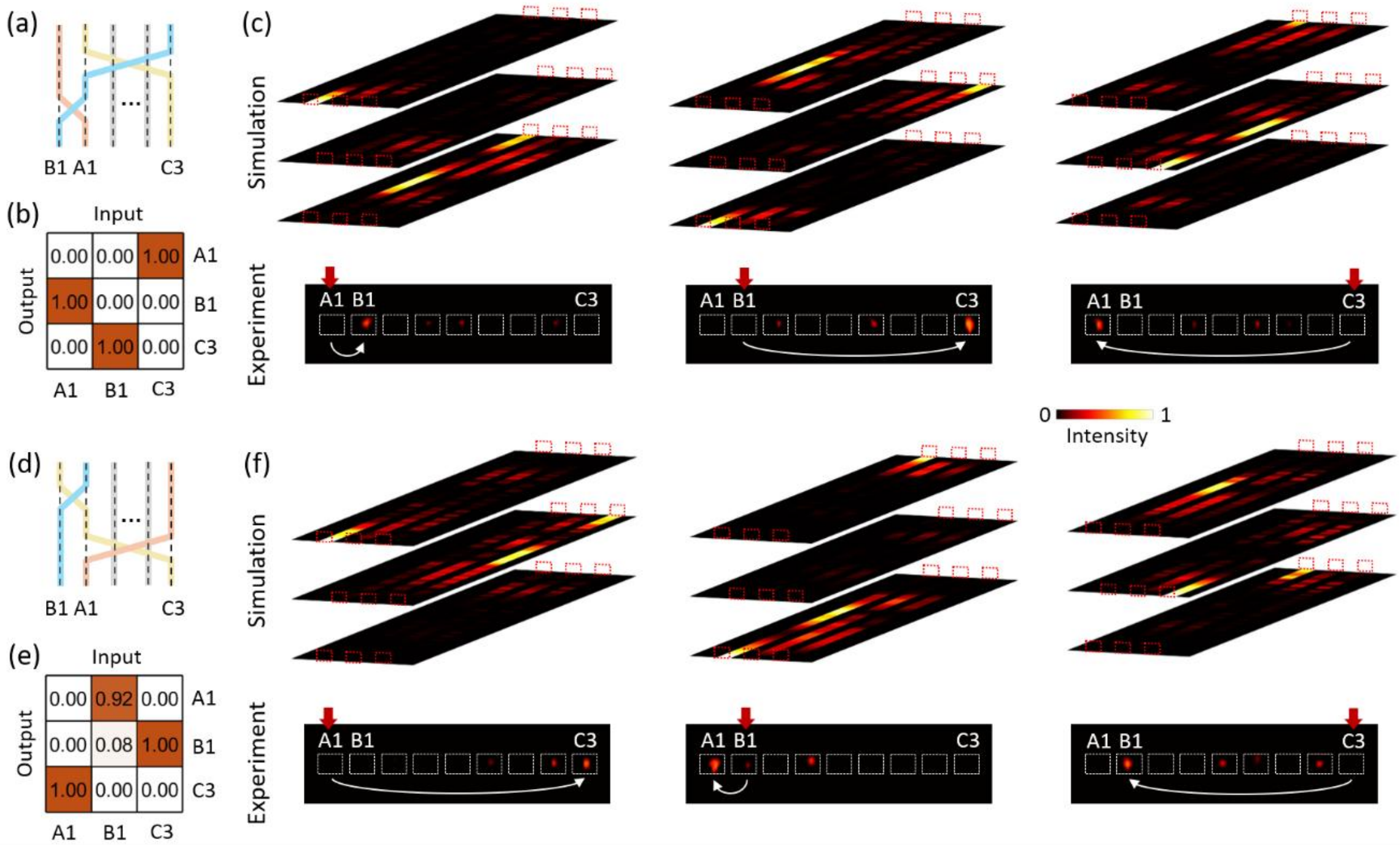

Fig. 4. Non-Abelian edge-state transfer via sequenced pumping channels. (a-c) Composite path in which the same-boundary channel is applied first, followed by the opposite-boundary channel. (a) Schematic of the cyclic permutation A1 → B1 → C3 → A1. (b) Measured 3×3 map among A1, B1 and C3. (c) Full-wave simulations (top) and measured output intensity distributions (bottom) for inputs at A1, B1 and C3 (red arrows). (d-f) Corresponding results for the reverse sequence, realizing the opposite cycle A1 → C3 → B1 → A1. Each elementary channel is 500μm long; a short connecting segment (5μm) is inserted between the two channels.

*Dissuasion.* In summary, we have demonstrated multi-channel non-Abelian edge pumping by adiabatically encircling multiple band degeneracies. The elementary building block is a three-band band-triangle in which two opposite-charge Weyl points and one zero-charge degeneracy are linked by global band connectivity. This connectivity enforces both opposite-boundary and same-boundary channels, allowing an edge mode to be selectively directed to either of the other two edge modes. Sequencing these channels produces non-Abelian braiding among topologically protected edge states on distinct boundaries. The evolution remains fully gapped, with the channels protected by topological winding rather than by maintained spectral degeneracies (see analysis on robustness in SM, Section 6). The same principle extends to an arbitrary number of channels by stacking additional band-triangles.

In the minimal three-channel system we experimentally demonstrate that sequencing the opposite-boundary and same-boundary channels yields distinct non-Abelian permutations. The same band-triangle connectivity further implies that, for larger $n$, a single adiabatic path enclosing an appropriate subset of degeneracies can redistribute multiple edge channels in one cycle, with the bulk bands acting as a shared relay (see SM, Section 5). This collective mechanism follows directly from the multi-gap spectral structure and does not require sequential pairwise swaps; its systematic realization for $n > 3$ remains a direction for future device implementation. The present work therefore establishes both the elementary non-Abelian edge operations and a scalable theoretical route toward multi-port edge routing protected by topological winding.

* songwange@nju.edu.cn
† taoli@nju.edu.cn
‡ shuzhang@hku.hk